# The Quantum Query Complexity of AC$^0$

Paul Beame[*]
Computer Science and Engineering
University of Washington
Seattle, WA 98195-2350
beame@cs.washington.edu

Widad Machmouchi[*]
Computer Science and Engineering
University of Washington
Seattle, WA 98195-2350
widad@cs.washington.edu

September 5, 2018

**Abstract**

We show that any quantum algorithm deciding whether an input function $f$ from $[n]$ to $[n]$ is 2-to-1 or almost 2-to-1 requires $\Theta(n)$ queries to $f$. The same lower bound holds for determining whether or not a function $f$ from $[2n-2]$ to $[n]$ is surjective. These results yield a nearly linear $\Omega(n/\log n)$ lower bound on the quantum query complexity of AC$^0$. The best previous lower bound known for any AC$^0$ function was the $\Omega((n/\log n)^{2/3})$ bound given by Aaronson and Shi's $\Omega(n^{2/3})$ lower bound for the element distinctness problem [1].

[*]Research supported by NSF grant CCF-0830626

# 1 Introduction

We study the quantum query complexity of Boolean functions that are computable with constant-depth polynomial-size circuits. The quantum query complexity of a function is the maximum number of queries a quantum algorithm needs to make to its input in order to evaluate the function (either with certainty, or with probability bounded away from 1/2 as we consider here). Both Boolean queries and $n$-ary queries can be natural depending on the context.

Starting with Grover's $O(\sqrt{n})$ quantum query algorithm for computing the OR of $n$ bits, and continuing through recent work on read-once formula evaluation [12, 7, 16, 6], quantum query algorithms have been shown to provide polynomial speed-ups over classical query algorithms for computing many total functions.

At the same time, three main techniques have been developed to derive lower bounds on quantum query complexity. The first two are the polynomial method of Beals et al. [8] and the adversary method introduced by Ambainis [2, 19]. The polynomial method uses the fact that the probability of a quantum algorithm succeeding is an approximating polynomial for the function that it is computing. On the other hand, Ambainis' adversary method is based on tracking the state of the quantum algorithm on "difficult" inputs. In a recent body of work [13, 17, 18, 15], the adversary method was extended to characterize the query complexity of Boolean functions and of functions defined on large alphabets. This can formulated as a version of the adversary method that includes negative weights [13] (positive weights do not help), as a certain measure on span programs computing the function [17], or as the optimum of a semi-definite program maximizing the spectral norm of the various *adversary matrices* of the function [18, 15]. An adversary matrix of a function is a real symmetric matrix indexed by pairs of inputs such that if the function evaluates to the same value on a pair of inputs, then the corresponding entry in the adversary matrix is set to 0. Our proof uses the original adversary method [2] and does not require the machinery developed in the subsequent work.

The methods above have been used to show that some fairly simple functions do not benefit from significant improvement in query complexity using quantum queries. For example, since the parity of $n$ bits requires degree $\Omega(n)$ to be approximated by a multivariate polynomial its quantum query complexity also must be linear. However, parity is still a somewhat complex function since it cannot even be approximated in AC$^0$, the class of decision problems solvable by Boolean circuits with unbounded fan-in, constant depth, and polynomial size. Can the simple functions in AC$^0$ always be computed using $O(n^\beta)$ queries for some $\beta < 1$? We show that this is not the case.

The quantum query complexity of a number of problems in AC$^0$ has been thoroughly investigated. The first quantum query lower bounds in [9, 8] showed that Grover's algorithm for the OR is asymptotically optimal. In [2], Ambainis used the adversary method to prove an $\Omega(\sqrt{n})$ lower bound on the query complexity of an AND of ORs. Motivated in part by questions of the security of classical cryptographic constructions with respect to quantum adversaries, the *collision problem* of determining whether a function is 1-to-1 or 2-to-1 on its domain has been one of the most celebrated examples in quantum query complexity. Aaronson and Shi [1] proved an $\Omega(n^{1/3})$ lower bound on the query complexity of the collision problem with range $3n/2$, which yields an $\Omega(n^{2/3})$ lower bound on the element distinctness problem with range $\Theta(n^2)$. The element distinctness problem is equivalent to the general case of testing whether a function is 1-to-1 (without the promise of being 2-to-1 in case that it is not 1-to-1). For the $r$-to-1 versus 1-to-1 problem, for integer $r \geq 2$, Aaronson and Shi also proved a lower bound of $\Omega((n/r)^{1/3})$ when the range size is $3n/2$. All of these problems can be computed in AC$^0$ and these query complexities are asymptotically optimal by results of Brassard et al. [10] and Ambainis [5]. Ambainis also was able to reduce the range size requirement for both lower bounds [3], as was Kutin [14] for the collision problem, independently, by different means.

The original adversary method has also been used to prove lower bounds for related *search* problems. In [11], the authors give an $\Omega(n)$ lower bound for the query complexities of a variety of problems where the goal is to produce a non-Boolean output, for example the index of some non-colliding input assuming that one exists.

In the original collision problem, the main question was whether or not the input function is 1-to-1, where at most $O(n^{1/3})$ queries suffice for the promise versions of the problem and at most $O(n^{2/3})$ queries suffice even in the general case. We show that if one is concerned with whether or not a function is precisely 2-to-1, the number of queries required is substantially larger. More precisely, we show an asymptotically tight linear lower bound for determining whether or not a function is either precisely 2-to-1 or *almost 2-to-1* in that the function is 2-to-1 except for exactly two inputs that are mapped 1-to-1. This also implies an $\Omega(n/\log n)$ lower bound on the Boolean quantum query complexity of $\mathsf{AC}^0$, thus substantially improving the previous best lower bound of $\Omega((n/\log n)^{2/3})$ given by the results of Aaronson and Shi.

Using the original adversary method developed by Ambainis [2] we prove the following theorem:

**Theorem 1.** *Let $n$ be a positive even integer, $\mathcal{F}$ be the set of functions $f$ from $[n]$ to $[n]$ that are either 2-to-1 or almost 2-to-1 and $\phi : \mathcal{F} \to \{0, 1\}$ be a Boolean function on $\mathcal{F}$ such that $\phi(f) = 1$ iff $f$ is 2-to-1. Then any quantum algorithm deciding $\phi$ requires $\Omega(n)$ queries.*

Noting that determining whether or not a function from $[2n - 2]$ to $[n]$ is surjective has as a special case the problem of determining whether a function is a 2-to-1 or almost 2-to-1 function, we obtain the following lower bound.

**Corollary 2.** *Let $n$ be a positive even integer and $\mathcal{G}$ be the set of functions from $[2n - 2]$ to $[n]$. Define the function* ONTO $: \mathcal{G} \to \{0, 1\}$ *by* ONTO$(f) = 1$ *if and only if $f$ is surjective. Then any quantum algorithm computing* ONTO *with probability at least $2/3$ requires $\Omega(n)$ queries.*

Since each value $f(i)$ can be encoded using $\log_2 n$ bits, the entire input is $n \log_2 n$ bits long. It is easy to see that determining whether such a function given by these bits is 2-to-1 or not is a simple $\mathsf{AC}^0$ problem. Since each query of one of the bits of $f(i)$ is weaker than querying all of $f(i)$, we immediately obtain lower bounds on the query complexity of $\mathsf{AC}^0$.

**Corollary 3.** *The quantum query complexity of $\mathsf{AC}^0$ is $\Omega(n/\log n)$.*

## 2 Lower bounds on quantum query complexity

Let $n$ be a positive even integer and $f : [n] \to [n']$ for $n' > n/2$ be a function. We say that $f$ is *2-to-1* if every point in the image either has exactly two pre-images or has no pre-images in $[n]$. We say $f$ is *almost 2-to-1* if $f$ is 2-to-1 on $n - 2$ points in the domain and 1-to-1 on the remaining 2 points, and every point has at most two pre-images.

We consider the following problem:

2-TO-1-VS-ALMOST-2-TO-1 Problem: Given a positive even integer $n$, integer $n' \geq n/2 + 1$, and a function $f : [n] \to [n']$ that is either 2-to-1 or almost 2-to-1, determine whether $f$ is 2-to-1.

The function $f : [n] \to [n']$ is given to us as a quantum oracle query which we can assume without loss of generality is accessed using a special register $i$ with answer register $j$. The oracle replaces a basis state $|i, j, \Psi\rangle$ by $|i, (f(i) + j) \mod n', \Psi\rangle$. The quantum query complexity of $f$ is the minimum number of queries needed by the quantum algorithm to correctly determine with probability at least $2/3$ whether $f$ is 2-to-1 or almost 2-to-1.

We will use the adversary method as developed by Ambainis to derive lower bounds on the quantum query complexity of Boolean functions.

**Theorem 4** (Ambainis [2]). *Let $\mathcal{F}$ be the set of functions $f$ from $[n]$ to $[n']$ and $\phi : \mathcal{F} \to \{0,1\}$ be a Boolean function. Let $A$ and $B$ be two subsets of $\mathcal{F}$ such that $\phi(f) \neq \phi(g)$ if $f \in A$ and $g \in B$. If there exists a relation $R \subset A \times B$ such that:*

1. *For every $f \in A$, there exist at least $m$ different functions $g \in B$ such that $(f, g) \in R$.*

2. *For every $g \in B$, there exist at least $m'$ different functions $f \in A$ such that $(f, g) \in R$.*

3. *For every $x \in [n]$ and $f \in A$, there exist at most $l$ different functions $g \in B$ such that $(f, g) \in R$ and $f(x) \neq g(x)$.*

4. *For every $x \in [n]$ and $g \in B$, there exist at most $l'$ different functions $f \in A$ such that $(f, g) \in R$ and $f(x) \neq g(x)$.*

*Then any quantum algorithm computing $\phi$ with probability at least $2/3$ requires $\Omega\left(\sqrt{\frac{mm'}{ll'}}\right)$ queries.*

Let $\mathcal{F}$ be the set of functions from $[n]$ to $[n']$ that are either 2-to-1 or almost 2-to-1 and let $\phi : \mathcal{F} \to \{0,1\}$ be a Boolean function such that $\phi(f) = 1$ if and only if $f$ is 2-to-1. We prove, using the adversary method [2], that the quantum query complexity of $\phi$ is $\Omega(n)$.

## 2.1 Quantum query lower bounds for the 2-TO-1-VS-ALMOST-2-TO-1 and ONTO functions

**Theorem 5.** *Let $n$ be a positive even integer, $n' \geq n/2 + 1$, and $\mathcal{F}$ be the set of functions $f$ from $[n]$ to $[n']$ that are either 2-to-1 or almost 2-to-1. Let $\phi : \mathcal{F} \to \{0,1\}$ be a Boolean function such that $\phi(f) = 1$ iff $f$ is 2-to-1. Then any quantum algorithm computing $\phi$ with probability at least $2/3$ requires $\Omega(n)$ queries.*

*Proof.* We use the adversary method. Let $A$ be the set of functions in $\mathcal{F}$ that are 2-to-1 and $B$ be the set of functions that are almost 2-to-1. Obviously, $\phi(A) = 1$ and $\phi(B) = 0$. We denote the distance between two functions $f$ and $g$ by
$$d(f, g) = |\{i \in [n] | f(i) \neq g(i), f, g \in \mathcal{F}\}|.$$
For a function $g \in B$, we denote the two input points in $[n]$ that $g$ maps injectively by $s(g) = \{s_1, s_2\}$ where $s_1 < s_2$. That is, $g(s_1) \neq g(s_2)$ and $g(i) \notin \{g(s_1), g(s_2)\}$ for all $i \notin s(g)$. Consider the following relation $R \subseteq A \times B$:
$$R = \{(f, g), f \in A, g \in B, d(f, g) = 2, f(s_1) = g(s_1) \text{ and } f(s_2) = g(s_2)\}.$$
In other words, the relation consists of pairs of functions in $A \times B$ that differ on exactly 2 points, neither of which is one of the injectively mapped points of the function in $B$.

**Claim 6.** *If $(f, g) \in R$ and $x, y \notin s(g) = \{s_1, s_2\}$ where $x \neq y$ are the points where $f$ and $g$ differ then $\{f(x), f(y)\} = \{g(s_1), g(s_2)\}$ and $g(x) = g(y)$.*

Since $f(s_1) = g(s_1) \neq g(s_2) = f(s_2)$, and $g(i) \notin \{g(s_1), g(s_2)\}$ for every $i \notin s(g)$, we must have $\{f(x), f(y)\} = \{g(s_1), g(s_2)\}$, otherwise $f$ will not pair any input with $s_1$ or $s_2$ and hence not be 2-to-1. Since $x, y \notin s(g)$, it must be that $x$ and $y$ are paired inputs of $g$. However, since $f$ is 2-to-1 and pairs up the four points $\{x, y, s_1, s_2\}$ all other points must be paired with each other by both $f$ and $g$ since $g$ agrees with $f$ on these other points. Hence the only points to which $x$ and $y$ can be paired by $g$ are each other, and therefore $g(x) = g(y)$. The claim follows.

We now check the four properties of $R$:

1. Let $f$ be a 2-to-1 function in $A$. Choose any two points $x, y \in [n]$ such that $f(x) \neq f(y)$. For each of the $n' - n/2$ points $z \in [n']$ not in the image of $f$ we can define an almost 2-to-1 function $g \in B$ such that $(f, g) \in R$ by having $g$ agree with $f$ on all but inputs $\{x, y\}$ and setting $g(x) = g(y) = z$. There are $n(n-2)/2$ choices of the pair of $x$ and $y$ and $n' - n/2$ choices of $z$, each of which produces a distinct $g$, for a total of $m = n(n-2)(2n'-n)/4$ distinct $g \in B$ with $(f, g) \in R$.

2. Let $g$ be an almost 2-to-1 function in $B$. Choose some ordered pair of inputs $(x, y) \notin s(g)$ such that $g(s_1) \neq g(x) = g(y) \neq g(s_2)$. Define a 2-to-1 function $f \in A$ such that $(f, g) \in R$ as follows: Let $f$ agree with $g$ on all inputs outside $\{x, y\}$. Define $f(x) = g(s_1)$ and $f(y) = g(s_2)$. There are $m' = n - 2$ choices of the ordered pair $(x, y)$ since there $n/2 - 1$ pairs of matched inputs for $g$ and 2 ways to order each pair.

3. Fix a function $f \in A$ and a point $x \in [n]$. Let $g \in B$ be such that $(f, g) \in R$ and $f(x) \neq g(x)$. Then $f$ and $g$ should differ an another point $y \in [n]$. By the claim, $f(x) \neq f(y)$ and $g(x) = g(y)$. There are precisely $n - 2$ choices of $y$. For each choice of $y$, the choice of $g$ is determined by the value $z = g(x) = g(y)$ which must not be in the image of $f$. (Indeed each such choice yields such a valid $g$.) Hence there are precisely $n' - n/2$ choices of $z$. So, in total there are precisely $l = (n-2)(2n'-n)/2$ choices of $g$ with $(f, g) \in R$ and $f(x) \neq g(x)$.

4. Fix a function $g \in B$ and a point $x \in [n]$. Let $f \in A$ be such that $(f, g) \in R$ and $f(x) \neq g(x)$. Then $f$ and $g$ should differ on another point $y \in [n]$. (We must have $x \notin s(g)$, otherwise there is no such $f$.) By the claim we must have $g(x) = g(y)$. Therefore there is only one such choice for $y$. Moreover by the claim we must have $\{f(x), f(y)\} = \{f(s_1), f(s_2)\}$ which yields precisely $l' = 2$ such functions $f$.

It follows that $\sqrt{\frac{m \cdot m'}{l \cdot l'}} = \sqrt{n(n-2)}/2$. Therefore the quantum query complexity of $\phi$ is $\Omega(n)$. $\square$

If $n' = n/2 + 1$ then observe that any almost 2-to-1 function is surjective but any 2-to-1 function is not. Therefore since $n = 2n' - 2$ we immediately obtain the following Corollary.

**Corollary 7.** *Let $n$ be a positive even integer and $\mathcal{G}$ be the set of functions from $[2n-2]$ to $[n]$. Define the function* ONTO $: \mathcal{G} \to \{0, 1\}$ *by* ONTO$(f) = 1$ *if and only if $f$ is surjective. Then any quantum algorithm computing* ONTO *with probability at least $2/3$ requires $\Omega(n)$ queries.*

## 2.2 A near optimal query lower bound for $AC^0$

**Corollary 8.** $AC^0$ *has worst-case quantum query complexity $\Omega(N/\log N)$ on $N$-bit functions.*

*Proof.* The lower bound in Theorem 5 is an asymptotically optimal $\Omega(n)$ for the 2-to-1 versus almost 2-to-1 problem. Even the total function $\phi$ which takes as input an arbitrary function $f : [n] \to [n]$ and determines whether or not it is 2-to-1 can be computed in $\mathsf{AC}^0$:

We represent the input $f : [n] \to [n]$ by $N = n \log_2 n$ bits, $f_{i\ell}$ where $f_{i\ell}$ represents the $\ell$-th bit of $f(i)$. The following is an explicit polynomial-size constant-depth formula that computes $\phi$:

$$\bigwedge_i \bigvee_{j \neq i} \bigwedge_{\ell=0}^{\log_2 n - 1} [(\neg f_{i\ell} \vee f_{j\ell}) \wedge (\neg f_{j\ell} \vee f_{i\ell})]$$

$$\wedge \bigwedge_{i \neq j \neq k \neq i} \bigvee_{\ell=0}^{\log_2 n - 1} [(\neg f_{i\ell} \wedge f_{j\ell}) \vee (\neg f_{i\ell} \wedge f_{k\ell}) \vee (\neg f_{j\ell} \wedge f_{i\ell}) \vee (\neg f_{j\ell} \wedge f_{k\ell}) \vee (\neg f_{k\ell} \wedge f_{i\ell}) \vee (\neg f_{k\ell} \wedge f_{j\ell})].$$

The first part of the formula computes whether for each input $i$ there is some input $j \neq i$ such that all bits of $f(i)$ and $f(j)$ agree and the second part determines whether for each triple of distinct inputs there is some bit on which some pair of $f(i), f(j), f(k)$ disagree. This unbounded fan-in formula clearly has size $O(n^3 \log n)$ and is depth 4 given its input literals. The lower bound of $\Omega(n)$ from Theorem 5 is $\Omega(N/\log N)$ as required.

We note that a similar result holds based on the ONTO function which has an even smaller $\mathsf{AC}^0$ formula:

$$\bigwedge_{j \in [n]} \bigvee_{i \in [2n-2]} \bigwedge_{\ell=0}^{\log_2 n - 1} f_{i\ell}^{j_\ell}$$

Where $j_\ell$ is the $\ell$-th bit of the binary encoding of $j$, and $x^{j_\ell}$ is $x$ if $j_\ell = 1$ and $\neg x$ if $j_\ell = 0$. This formula is only depth 3 and size $O(n^2 \log n)$. □

It is interesting to note the importance of the domain size in this problem. Testing surjectivity for functions from $[n]$ to $[n]$ is equivalent to testing whether two elements from the domain are mapped to the same point in the range and thus has query complexity $\Theta(n^{2/3})$, given by the element distinctness problem [1].

## 3 Discussion

While we have shown that there is a function $\phi : [n]^n \to \{0,1\}$ in $\mathsf{AC}^0$ requiring linear quantum query complexity, when this is encoded using Boolean inputs it requires $\Omega(n \log n)$ bits and thus the quantum query lower bound is lower than the number of input bits by an $O(\log n)$ factor. (Note that this lower bound holds even when the query algorithm is able to read the group of $\log_2 n$ bits surrounding each query bit at no extra cost.) This does rule out a polynomial speed-up but it would be nice to obtain a linear lower bound for the Boolean case.

Also, our results do nothing to close the gap in the lower bound on the approximate degree of $\mathsf{AC}^0$ functions. In [4], Ambainis used a standard recursive construction to obtain a Boolean function with quantum query complexity larger than its approximate degree by a small polynomial amount, thus giving a separation between query complexity and approximate degree. Determining the approximate degree of the 2-TO-1-VS-ALMOST-2-TO-1 or ONTO functions, would yield either a much larger degree lower bound for $\mathsf{AC}^0$ or a much larger gap between approximate degree and quantum query complexity than is currently known.

# Acknowledgements

We thank Parikshit Gopalan for suggesting the problem of the approximate degree of AC$^0$ which motivated this work. We also thank Scott Aaronson and Dave Bacon for helpful pointers.